\documentclass[11pt,a4paper]{article}
\usepackage[margin=2.5cm]{geometry}

\usepackage[T1]{fontenc}
\usepackage{lmodern}
\usepackage[utf8]{inputenc}

\usepackage{amsmath, amsfonts, amssymb, amsthm}
\usepackage{mathtools}
\usepackage{bm}              
\usepackage{nicefrac, xfrac}
\usepackage[scr=rsfso]{mathalfa}

\usepackage{siunitx}
\providecommand{\angstrom}{\text{Å}}
\DeclareSIUnit\angstrom{\angstrom}

\usepackage{graphicx}
\usepackage{booktabs}
\usepackage{array}
\usepackage{multirow}
\usepackage{multicol}
\usepackage{tabularx}
\usepackage{float}
\usepackage{subcaption}
\usepackage{enumitem}
\usepackage{soul}

\graphicspath{{figures/}}

\usepackage{authblk}
\usepackage{hyperref}

\title{Micrograph B-factor estimation via Wilson statistics}

\author[a,*]{Ayelet Heimowitz}
\author[b]{Amit Singer}

\affil[a]{Department of Electrical and Electronics Engineering, Ariel University, Israel}
\affil[b]{Department of Mathematics, Program for Applied and Computational Mathematics, Princeton University, Princeton, USA}

\date{} 

\begin{document}

  \maketitle
  \begin{abstract}
  \noindent The single-particle cryo-electron microscopy imaging process is known to cause exponential decay in the amplitude and power spectra of proteins. This decay is modeled by the B factor and is currently corrected  as a post-processing stage of the 3D reconstruction. As a consequence,  all experimental images in the dataset are assumed to share the same B factor. In this paper, we suggest a novel approach wherein the B factor is determined and corrected as a preprocessing step to the reconstruction. We show such an approach can improve reconstruction resolution. 
  \nolinebreak
  \end{abstract}

\section{Introduction}
Together with X-ray crystallography and NMR spectroscopy, cryo electron-microscopy (cryo-EM) is one of the leading methods for determination of the structure of macromolecules at near atomic resolutions. In this technique, a frozen-hydrated biological specimen is imaged in an electron microscope, resulting in multiple 2-D projections of the macromolecule. These projections are then used for structure determination \cite{frank1996, singer2020}.

Unfortunately, the imaging process introduces many distortions to the projections. First, during the imaging process a
high-energy electron beam is shot through the ice and the macromolecules frozen within, causing radiation damage. The damage
begins with an attenuation of the signal in the high frequencies and may completely destroy the sample, depending on the electron
dose~\cite{rosenthal2003}. The electron dose is therefore limited, leading to high levels of shot noise~\cite{henderson1995, jensen2001, bai2013}. 

In addition to the radiation damage and shot noise, further issues are caused by specimen motion. During the imaging process,
a drum-like motion is experienced by the carbon support film, leading to translation and rotation of the macromolecules embedded
in the ice~\cite{brilot2012}. This motion leads to blurring of the tomographic projections, causing further attenuation of the
high frequency information. Beyond this, the issue of heterogeneity of the macromolecule\footnote{As the macromolecules are not fully rigid.} leads to additional decay of the structure factor
amplitudes~\cite{jensen2001}.

It follows that the power spectrum of the tomographic projections experiences exponential decay due to the imaging process. 
It is common to model this exponential decay as determined by the B factor (temperature factor). Currently, after resolving the 3-D structure of the macromolecule, a B-factor correction is performed by flattening the power spectrum of the reconstructed map, thereby correcting for the decay \cite{fernandez2008, vilas2020reexamining}.\footnote{It is important to note that flattening the power spectrum of the reconstructed map will not only correct the imaging-driven exponential decay that was detailed above, it will  also correct an additional exponential decay,  modeled by an additional  B factor. This second decay occurs due to the reconstruction procedure itself (for example, imperfect estimation of the viewing angles) and will still need to be estimated and corrected at the end of the reconstruction process. Crucially, as this paper deals with a stage in the cryo-EM pipeline where this computation-based exponential decay has not yet occurred, we do not refer to it, and therefore  use the term B factor for the imaging-driven exponential decay alone.}

In this paper we suggest that, rather than correcting the B-factor at the end of the reconstruction process, it can be estimated and  corrected as a preprocessing step on a  per-micrograph basis. In this manner, the correction can be done directly after CTF estimation, prior to 3-D reconstruction.

To this end, we analyze the power spectrum of the cryo-EM experimental images and compare to the true power spectra of
proteins, as detailed from Guinier law and Wilson statistics. We use this comparison to suggest a method for B-factor determination. 
We show in the Experimental Section that flattening the mid-high frequency information can lead to improved resolution reconstructions.

\section{Methods}
\subsection{Power Spectrum of proteins: ideal vs. imaging effects}

In this Section, we analyze the power spectrum of the protein and the power spectrum of the micrograph. We then formulate
the connection between the power spectrum of a micrograph and the ideal power spectrum of the macromolecule projections. Lastly, we will use this connection to estimate the B-factor for each  micrograph.

\subsubsection{The PSD of macromolecules (proteins)}

 Wilson statistics and Guinier law for the decay of the power spectrum of molecules provide us with a characterization  
for the power spectral densities (PSD) of proteins. According to Guinier law, the PSD of a protein will exhibit a steep exponential decay at low frequencies \cite{rosenthal2003, singer2021}. This exponential decay is influenced by the radius of gyration
$\text{R}_{\text{g}}$ of the protein. The PSD of proteins can therefore be modeled, at the low frequencies, as follows~\cite{rosenthal2003}
\begin{equation}
    \text{P}_{\text{mol}} \left( \mathbf{g} \right) = \left(0.42N\right)^2 \mathrm{e}^{-\frac{\text{R}_\text{g}^2}{3} \vert \mathbf{g} \vert^2},
    \label{equ:ideal_psd}
\end{equation}
where $\mathbf{g} = \left(g_1,g_2,g_3 \right)$ is the spatial frequency, 0.42 is the solvent factor and N is the mass of the molecule. While Guinier law defines the behavior of the PSD at low frequencies, Wilson statistics states that this decay is followed by a (nearly) flat slope at mid-high frequencies  or, more precisely, the structure factor of a carbon atom~\cite{Gilles2022}.

\subsubsection{Imaging effects: the PSD of micrographs}

For the purpose of formulating the PSD of a micrograph, we  use the weak-phase object approximation, under which the image formation model of a micrograph can be described as follows~\cite{frank1996, thon1971, mindell2003} 
\begin{equation}
    y = h_{\phi}*x+e,
    \label{equ:ifm}
\end{equation}
where $x$ is an image containing the clean tomographic projections, $h_{\phi}$ is the point spread function which is determined by characteristics of the objective lens, and $e$ is additive noise~\cite{frank1996, heimowitz2020}. Assuming the micrograph is not overcrowded, its PSD  is 
\begin{equation}
    S_y \left( \mathbf{k} \right) = \vert H_{\phi}\left( \mathbf{k} \right) \vert^2 S_x \left( \mathbf{k} \right) +S_e\left( \mathbf{k} \right),
    \label{equ:mic_psd}
\end{equation}
where $S_x$ and $S_e$ are the PSD of $x$ and the noise, correspondingly, and $ \mathbf{k} = \left(k_1, k_2 \right)$ is the spatial frequency. The
contrast transfer function $H_{\phi}\left( \mathbf{k} \right)$  is the Fourier transform of the point spread function. This function has a simple expression in
the polar coordinates of the spatial frequency. It is modeled as a sine function whose argument depends on the spatial frequency 
and several parameters of the objective lens of the microscope~\cite{frank1996}. Specifically,
\begin{equation}
    H_{\phi}\left( \mathbf{k} \right) = - \sin{\left( \frac{1}{2} \pi \lambda^2 \vert k \vert^2 2 \Delta f \left( \alpha \right) - \frac{1}{2} \pi \lambda^3 \vert k \vert^4 C_s + w \right)},
\end{equation}
where $\lambda$ is the electron wavelength, $C_s$ is the spherical aberration, $w$ is the amplitude contrast and $\Delta f \left( \alpha \right)$ is determined by the defocus of the objective lens.

The power spectrum of the noise $S_e$, also known as the background, is typically assumed to be a smooth function and can be estimated in various methods~\cite{mindell2003, heimowitz2020}. After the background is estimated, it can be subtracted from the PSD of the micrograph, resulting in the background-subtracted PSD
\begin{equation}
    S_y \left( \mathbf{k} \right) - S_e \left( \mathbf{k} \right) = \vert H_{\phi}\left( \mathbf{k} \right) \vert^2 S_x \left( \mathbf{k} \right).
    \label{equ:mic_psd}
\end{equation}

\begin{figure}[h]
    \centering
    \begin{subfigure}{0.23\textwidth}
        \includegraphics[width=\linewidth]{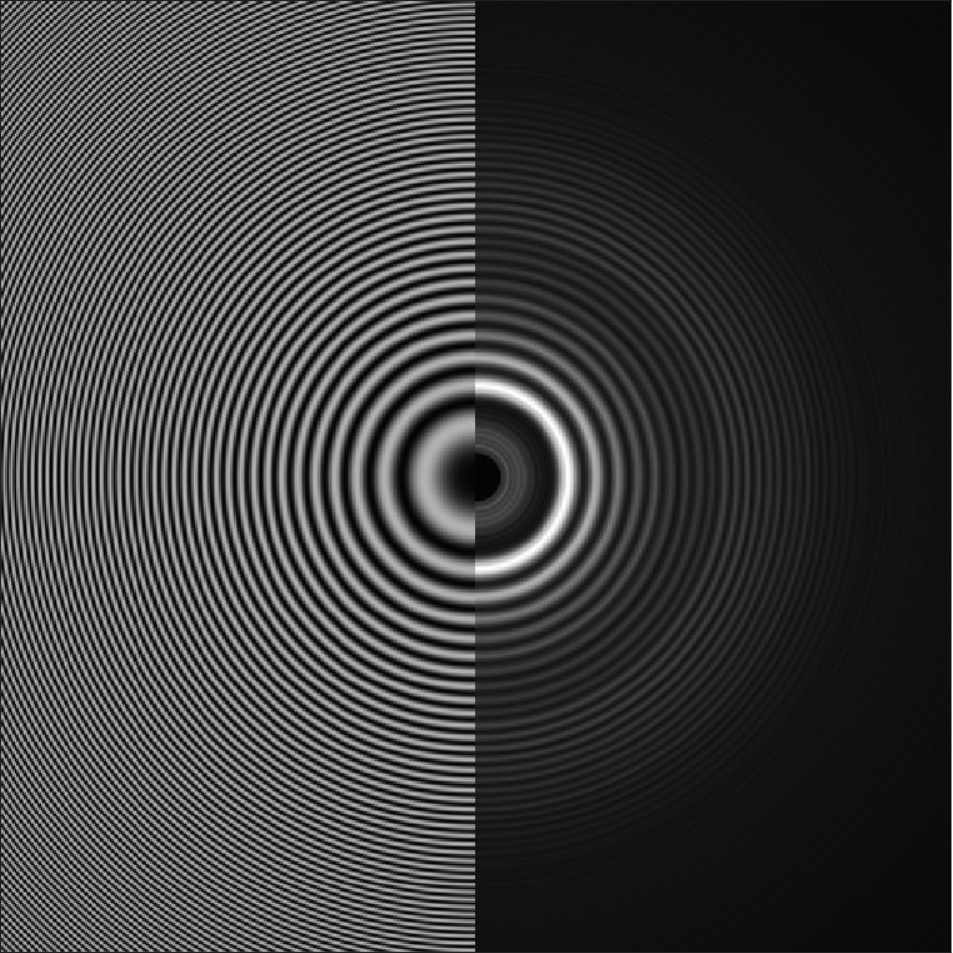}
        \caption{}
        \label{fig:ctf1}
    \end{subfigure}
    \hspace{1cm}
    \begin{subfigure}{0.5\textwidth}
        \centering
        \includegraphics[width=\linewidth]{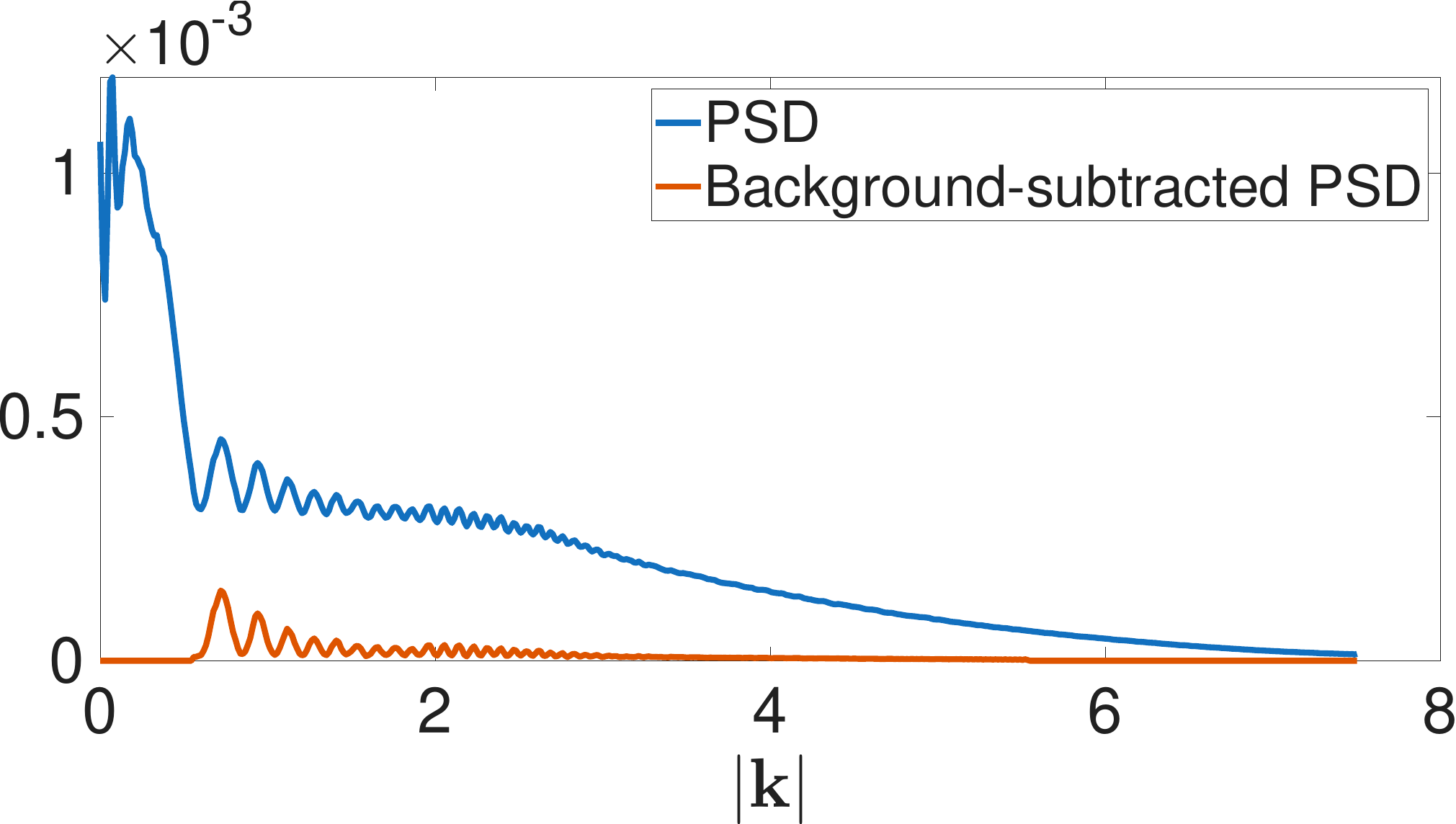}
        \caption{} 
        \label{fig:psd_1d}
    \end{subfigure}
    
    \caption{Background-subtracted PSD ($S_y \left( \mathbf{k} \right) - S_e\left( \mathbf{k} \right)$) of a micrograph from EMPIAR-10420 dataset. (a) The left half of the image is the CTF of the micrograph, while the right half contains the estimated background subtracted PSD. (b) The 1-D radial profile of the PSD of the micrograph as well as the 1-D radial profile of the background-subtracted PSD. We note the units of $\vert k \vert$ are $1/$nm.}
    \label{fig:psd}
\end{figure}

We present in Figure~\ref{fig:psd} the background-subtracted PSD of a micrograph from the EMPIAR-10420 dataset~\cite{iudin2023empiar, tan2020}.\footnote{We set the very low frequencies of the  background-subtracted PSD to $0$ to avoid artifacts from background subtraction. Note that this frequency range is never used for B factor estimation.} We note that the PSD exhibits (a) Thon rings~\cite{thon1971}, which are the result of the multiplication by the squared CTF in~\eqref{equ:mic_psd}, (b) a steep exponential decay in low frequencies
and (c) a moderate decay in mid-high frequencies.

The steep exponential decay in low frequencies is due to Guinier law~\eqref{equ:ideal_psd}, as well as  imaging issues such as radiation damage, specimen motion and heterogeneity. The moderate decay in mid-high frequencies is due to imaging issues.  

The variation of the background-subtracted PSD of the micrograph from  Wilson statistics is often modeled as a multiplication between $\vert H_{\phi}\left(\mathbf{k}\right)\vert$ and the  following 
exponentially decaying envelope function~\cite{fernandez2008, jensen2001}
\begin{equation}
    E_B \left( \mathbf{k} \right)^2 = \mathrm{e}^{-\frac{\mathrm{B}}{2} \vert \mathbf{k} \vert^2},
    \label{equ:b}
\end{equation}
where B is the B factor, measured in \r{A}$^2$ (or $\mathrm{nm}^2$) and $\mathbf{k}$ contains the mid-high frequency range. The PSD of the micrograph~\eqref{equ:mic_psd}
can therefore be rewritten as
\begin{equation}
     S_y \left( \mathbf{k} \right) - S_e \left( \mathbf{k} \right) = \vert H_{\phi}\left( \mathbf{k} \right) \vert^2 S_{\tilde{x}} \left( \mathbf{k} \right) E_{\textrm{B}}\left(\mathbf{k}\right)^2 =  \vert H_{\phi}\left( \mathbf{k} \right) \vert^2 S_{\tilde{x}} \left( \mathbf{k} \right) \mathrm{e}^{-\frac{\mathrm{B}}{2} \vert \mathbf{k} \vert^2},
    \label{equ:mic_psd_b}
\end{equation}
where $S_{\tilde{x}}$ is the ideal PSD of the protein.

It follows that the exponential decay in the PSD of the micrograph changes between low frequencies and high frequencies. We can
therefore divide the frequency range into three domains. The first domain contains the low frequencies. In this domain the PSD of the protein is determined by Guinier law~\eqref{equ:ideal_psd} and the PSD of the micrograph exhibits an exponential decay due to both~\eqref{equ:ideal_psd}
and~\eqref{equ:b}. The upper bound on this range is $s^{-1}$ where $s$ is the size of the molecule. The second frequency domain consists of mid-high frequencies, typically in the range $10 \text{\r{A}} - 5 \text{\r{A}}$. In this domain the PSD of the molecule is approximately flat in consistence with Wilson statistics. The exponential decay exhibited in this frequency range is due to the imaging process and is of the form~\eqref{equ:b}. The third range is the high frequencies. At these frequencies the PSD of the molecule is still (mostly)  approximately flat, however, the loss of signal in the micrograph is rather extreme and, therefore, its PSD will be dominated by noise fluctuations. Furthermore, this range does exhibit a variation from Wilson statistics, specifically a spike at $3.7$\AA,  due to the effects of the amorphous ice~\cite{McMullan2015}. This third range should therefore be ignored. 

\subsection{Recovering the B-factor}
We suggest that the B factor can be recovered from the mid-high frequency range. At this range~\eqref{equ:mic_psd_b} can be rewritten as follows,
\begin{equation}
    \frac{S_y \left( \mathbf{k} \right) - S_e \left( \mathbf{k} \right) }{\vert H_{\phi} \left( \mathbf{}{k} \right) \vert^2 S_{\tilde{x}} \left( \mathbf{k} \right)}= 
     \mathrm{e}^{-\frac{B}{2} \vert \mathbf{k} \vert^2}, \quad 0.1 \le \vert \mathbf{k} \vert \le 0.2.
    \label{equ:mic_all}
\end{equation}

Since the CTF is a sine function, we focus on the set of frequencies within the mid-high frequency range where $H_{\phi} \left( \mathbf{k} \right)$ reaches a maximum or a minimum. At these frequencies $\vert H_{\phi} \left( \mathbf{k} \right) \vert$ simply equals $1$ and can therefore be ignored. Formally, we define $G_m$ as
the set of spatial frequencies in the mid-high frequency range where the CTF reaches a maximum or a minimum. We can now
reformulate~\eqref{equ:mic_all} as follows,
\begin{equation}
    \frac{S_y \left( \mathbf{k} \right) - S_e \left( \mathbf{k} \right)}{S_{\tilde{x}} \left( \mathbf{k} \right)} = \mathrm{e}^{-\frac{B}{2} \vert \mathbf{k} \vert^2}, \qquad k \in G_m.
    \label{equ:psd_at_opt}
\end{equation}
While $S_{\tilde{x}}$ as a whole is unknown, Wilson statistics show us that at the mid-high frequency range it is approximately flat. It follows that, in  this frequency range, $S_{\tilde{x}}$ can be approximated as a line with $0$ slope, or, more accurately, we can approximate it by the square structure factor of carbon, computed according to~\cite{peng1996},
\begin{equation}
    S_{\tilde{x}} \left( \mathbf{k} \right) = 0.0489\mathrm{e}^{-0.1140k^2} + 0.2091\mathrm{e}^{-1.0825k^2} + 0.7537\mathrm{e}^{-5.4281 k^2} + 1.1420\mathrm{e}^{-17.8811k^2} + 0.3555\mathrm{e}^{-51.1341k^2}.
    \label{equ:sf}
\end{equation}
   
It follows from~\eqref{equ:psd_at_opt} that displaying a Guinier plot of the natural logarithm of the background subtracted PSD of the micrograph $ S_y \left( \mathbf{k} \right) - S_e \left( \mathbf{k} \right)$, divided by the PSD of carbon, at the frequencies in $G_m$ will result in a line with a slope of $-\frac{B}{2}$.

\begin{figure}[h!]
    \centering
    \begin{subfigure}{0.4\textwidth}
        \centering
        \includegraphics[width=\linewidth]{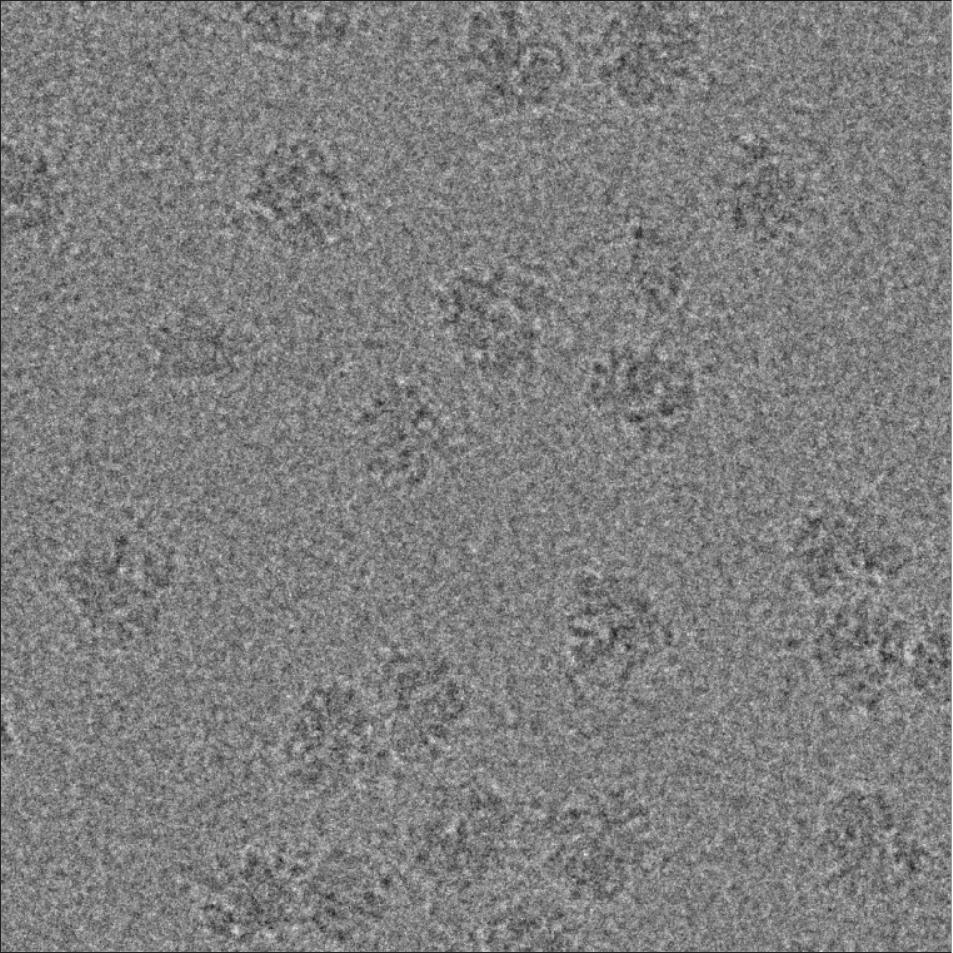}
        \caption{} 
        \label{fig:micro}
    \end{subfigure}
    \hfill
    \begin{subfigure}{0.4\textwidth}
        \includegraphics[width=\linewidth]{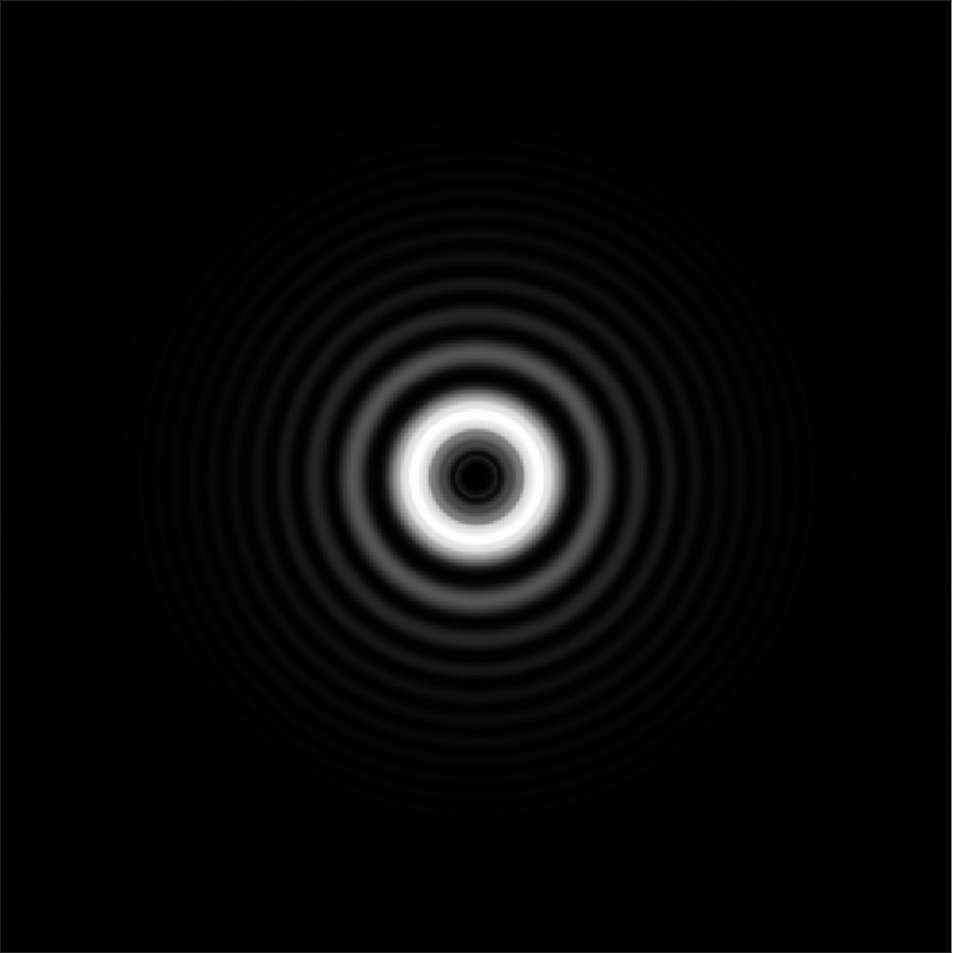}
        \caption{}
        \label{fig:2dpsd}
    \end{subfigure}
    \begin{subfigure}{0.48\textwidth}
        \centering
        \includegraphics[width=\linewidth]{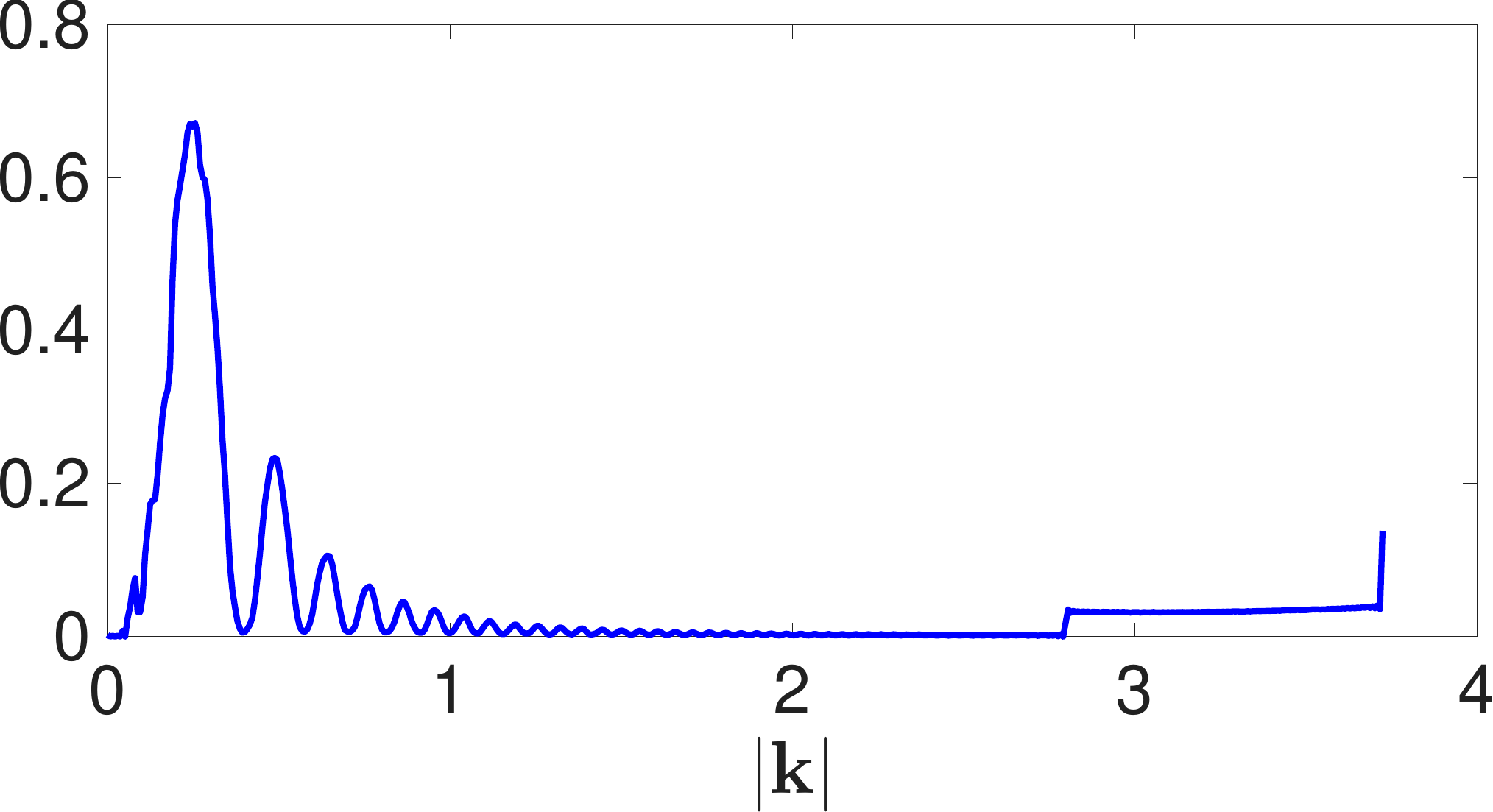}
        \caption{} 
        \label{fig:psd1d}
    \end{subfigure}
    \begin{subfigure}{0.48\textwidth}
        \includegraphics[width=\linewidth]{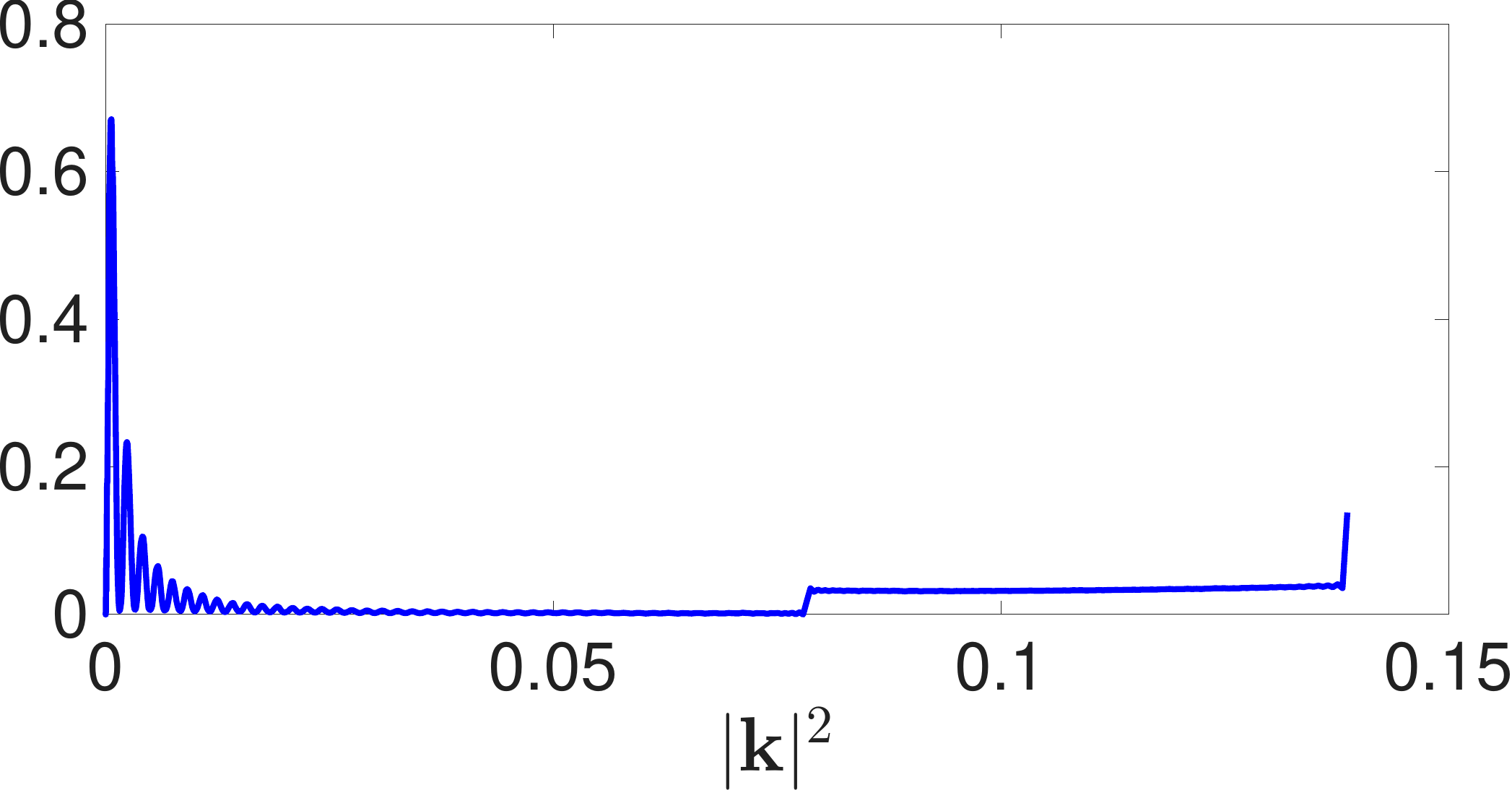}
        \caption{}
        \label{fig:guinier}
    \end{subfigure}
    \begin{subfigure}{0.48\textwidth}
        \centering
        \includegraphics[width=\linewidth]{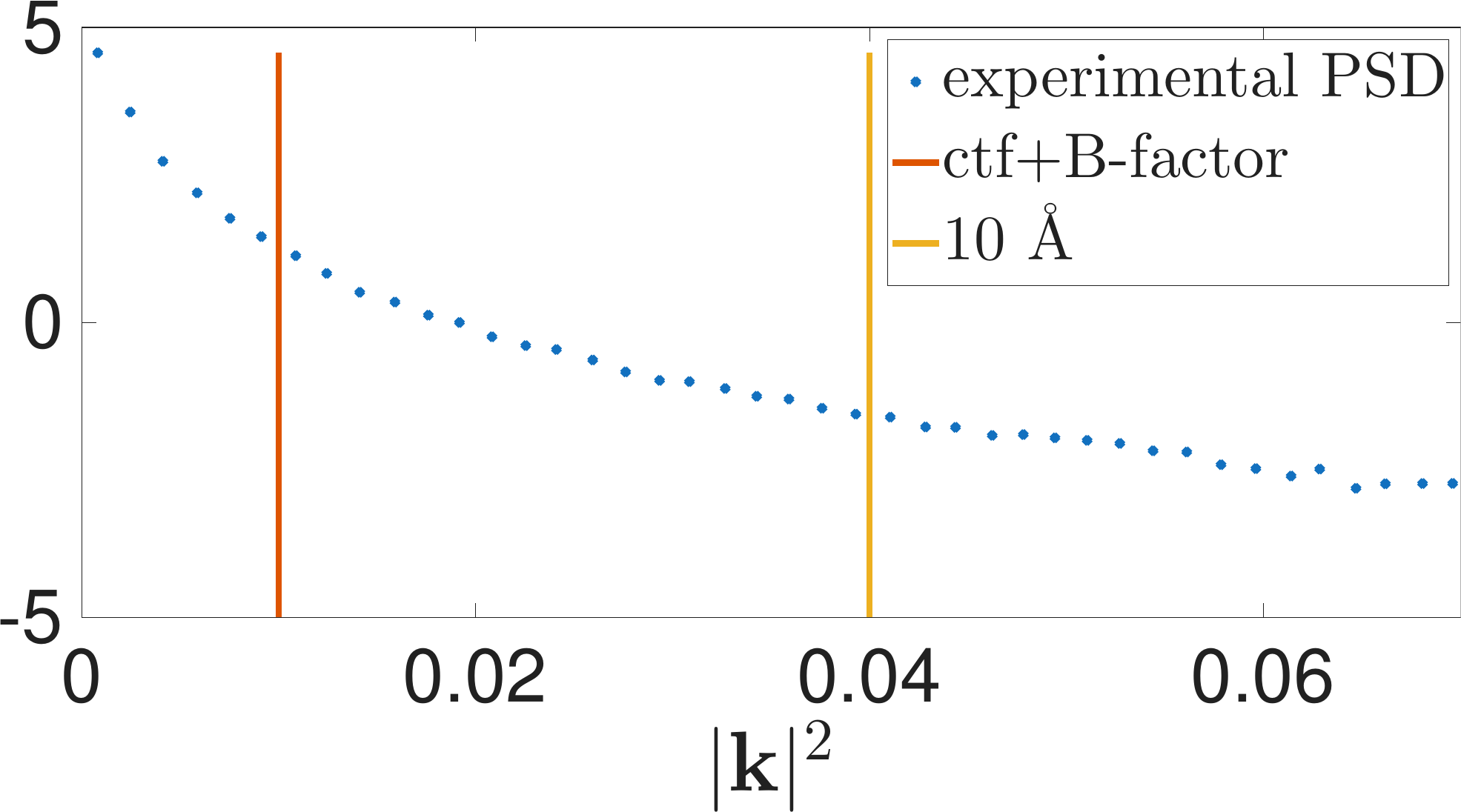}
        \caption{} 
        \label{fig:log}
    \end{subfigure}
    \begin{subfigure}{0.48\textwidth}
        \centering
        \includegraphics[width=\linewidth]{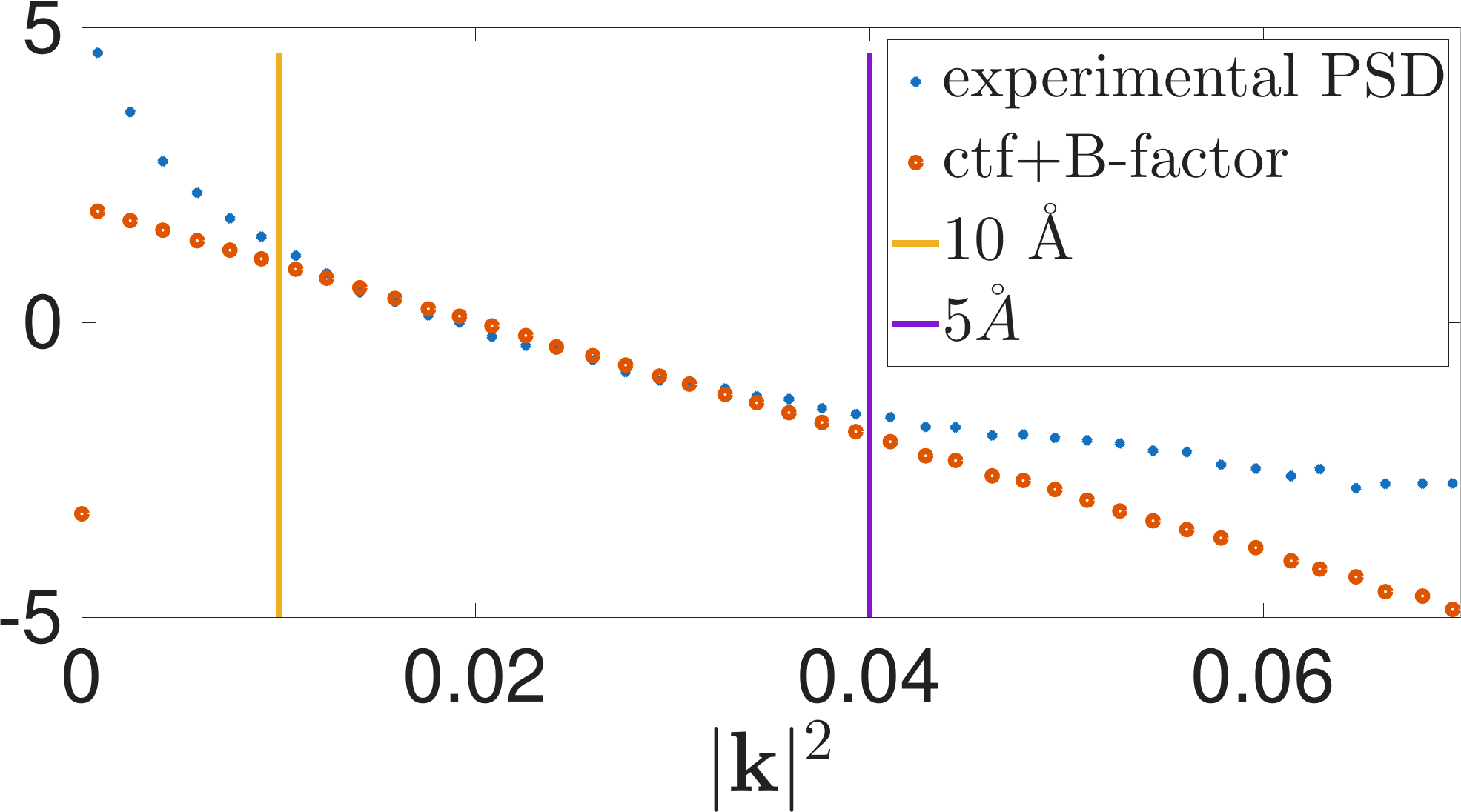}
        \caption{} 
        \label{fig:estimate}
    \end{subfigure}
    \caption{Outline of B-factor estimation. (a) A window extracted from a micrograph of the EMPIAR-10028 dataset. (b)  The background-subtracted PSD of the micrograph. (c) The radial profile of the background-subtracted PSD of the micrograph after division by the carbon structure factor~\eqref{equ:sf}. (d) Guinier plot of the 1D background subtracted PSD after division by the carbon structure factor. (e) Guinier plot of the logarithm of the background subtracted PSD  of the micrograph (normalized for convenience), divided by the PSD of carbon, at the frequencies in $G_m$. (f) Tile (e) overlaid with the estimated exponential decay determined by the B-factor (normalized for convenience), at the frequencies $G_m$ where the squared CTF reaches its maxima.}
    \label{fig:bfactor}
\end{figure}

We present in Figure~\ref{fig:bfactor} a visualization of our method. We present a section of a micrograph in Figure~\ref{fig:micro}. We present its 2D and 1D background subtracted PSD in Figures~\ref{fig:2dpsd} and~\ref{fig:psd1d}, respectively. We present the Guinier plot of the 1D background subtracted PSD in Figure~\ref{fig:guinier}. Furthermore, we present the Guinier plot at frequencies $G_m$ in Figure~\ref{fig:log} and the line with slope $-\frac{B}{2}$ in Figure~\ref{fig:estimate}. 

\section{Results}

We tested our method on the real-world datasets  EMPIAR-10017~\cite{iudin2023empiar, scheres2015}, EMPIAR-10028~\cite{wong2014cryo} and EMPIAR-10420~\cite{tan2020}. Specifically, for each dataset we used RELION to compute two 3D reconstructions. In the first reconstruction we used the traditional pipeline, where the B factor is computed globally as a post processing step to the reconstruction. In the second reconstruction we corrected the decay due to the B factor in each micrograph before beginning the reconstruction pipeline. Our method is shown to improve reconstruction resolution. 

\subsection{Beta-galactosidase}
\label{subsec:beta-gal}
The EMPIAR-10017~\cite{iudin2023empiar, scheres2015} dataset contains 84 Beta-galactosidase micrographs along with the coordinates
of manually selected projections. We used RELION~\cite{scheres2012relion, zivanov2018new} to create two reconstructions from these coordinates.\footnote{Coordinates were only used to create a reconstruction. They are not used for B factor estimation.} The first reconstruction was done using the original micrographs with B factor correction done post reconstruction, while the second reconstruction was done using micrographs corrected for the exponential decay caused by the B factor. 
Each reconstruction consisted of outlier detection via 2D and 3D-classification, initial model building, refinement and postprocessing.

The B factor identified by our method ranges between $48{\text{\AA}}^2$ and $161{\text{\AA}}^2$, with an average of $95\text{\AA}^2$ and standard deviation of $27.8{\text{\AA}}^2$. As expected, there is variability between the B factor in different micrographs, leading to the conclusion that computing the B factor individually for each micrograph can lead to improved results. Indeed, we obtained a 0.2\r{A} improvement to resolution when using micrographs modified for the exponential decay of the 
envelope function as opposed to the original micrographs. Results are summarized in Table~\ref{tab:projection_comparison_stacked}.

The B factor computed during RELION's post-processing task is $212\text{\AA}^2$.\footnote{The B-factor was expected to be larger when computed during the Post-processing task of RELION since it now accounts for decay that stems from reconstruction computations rather than image acquisition alone.}

\subsection{Plasmodium falciparum 80S ribosome}
The EMPIAR-10028~\cite{iudin2023empiar, wong2014cryo} dataset contains 1081 Plasmodium falciparum 80S ribosome micrographs along with picked particle coordinates. We once again followed the pipeline detailed in Section~\ref{subsec:beta-gal}. Once again, correcting the B factor prior to the reconstruction pipeline leads to a $0.1$\r{A}   improvement in resolution.

The B factor identified by our method ranges between\footnote{There is a single micrograph in the dataset where the B factor is severely underestimated at $13\text{\AA}^2$. The remaining $1080$ micrographs contain a B factor in the reported range.} $72\text{\AA}^2$ and $167\text{\AA}^2$, with an average of $120\text{\AA}^2$ and standard deviation of $12.98{\text{\AA}}^2$.

Reconstruction results are summarized in Table~\ref{tab:projection_comparison_stacked}.

\subsection{Arabinosyltransferase EmbB}
The EMPIAR-10420~\cite{iudin2023empiar, tan2020} entry consists of two datasets. The first was acquired using the Falcon III direct detector and the second was acquired using the K2 direct detector. We focus on the first of these datasets. The dataset includes $19,755$ projection coordinates. These coordinates are supplied only after outlier detection done in both RELION and CryoSPARC~\cite{Punjani2017cryosparc}. We performed one reconstruction using the above $19,755$ coordinates and correcting B factor post reconstruction and, additionally, corrected the B factor of all the micrographs and performed a second reconstruction using the same coordinates. A third reconstruction was done from the B factor-corrected micrographs, where an additional 2D classification was performed using RELION, discarding approximately $7000$ projections.

Notably, the first two reconstructions achieved the same resolution of $6.4$\text{\AA}, while the third reconstruction achieved a resolution of $5.2$\text{\AA}. It seems that emphasizing the mid-high frequency range allowed the 2D classification process to distinguish and discard projections with damaged high resolution content. 

We note that all three reconstructions were done after downsampling the projections by a factor of $1.5$ as suggested in~\cite{tan2020}. Unlike~\cite{tan2020} we did not do per-particle CTF estimation or particle polishing for improved reconstruction resolution.

The B factor identified by our method ranges between $33\text{\AA}^2$ and $149\text{\AA}^2$, with an average of $83\text{\AA}^2$ and standard deviation of $17.3{\text{\AA}}^2$.

\section{Conclusion}

In this paper we have presented a novel method for B factor 
estimation. We suggest that these parameters can be estimated as a preprocessing step, directly from each micrograph.  In this manner, we correct the exponential decay caused by the imaging process prior to 2D classification. We tested our modified pipeline on three real-world datasets, EMPIAR-10017, EMPIAR-10028 and EMPIAR-10420 and show that B factor correction as a preprocessing step can improve reconstruction resolution.

Following our experimental verification we attempted to identify the part of the imaging process that causes variations in the B factor across micrographs within the same dataset, as these micrographs do share exposure time and electron dose. We did not find a direct connection to defocus or astigmatism, however, the drum-like motion experienced during image acquisition is different between micrographs, and the quality of motion correction can therefore vary somewhat between different micrographs. It is therefore likely the variation in B factor is linked to the success of the motion correction procedure.

\begin{table}
    \centering
    \renewcommand{\arraystretch}{1.2} 
    \begin{tabular}{lccccc}
        \toprule
        \textbf{Metric} & \textbf{Original} & \textbf{Ours} & Mean B factor  & B factor  \\ & & & (Micrographs) & (RELION)\\
        
        \midrule
        
        \multicolumn{3}{l}{\textbf{EMPIAR-10017}} \\ 
        \midrule
        \# Projections used in reconstruction & 29,814 & 25,933 \\ 
        Resolution & 4.4\text{\AA} & 4.2\text{\AA} & $95\text{\AA}^2$ & $212\text{\AA}^2$\\ 
        
        \bottomrule
        \multicolumn{3}{l}{\textbf{EMPIAR-10028}} \\ 
        \midrule
        \# Projections used in reconstruction & 117,465 & 139,359 \\ 
        Resolution & 3.7\text{\AA} & 3.6\text{\AA} & $120\text{\AA}^2$ & $143\text{\AA}^2$ \\ 
        \bottomrule

        \multicolumn{3}{l}{\textbf{EMPIAR-10420}} \\ 
        \midrule
        \# Projections used in reconstruction & 19,755 & 12,456 \\ 
        Resolution & 6.4\text{\AA} & 5.2\text{\AA} & $83\text{\AA}^2$ & $178\text{\AA}^2$\\ 
        \bottomrule
    \end{tabular}
    \caption{Final reconstruction metrics. }
    \label{tab:projection_comparison_stacked}
\end{table}

\section*{Data Availability}
All experimental cryo-EM data used in this study are publicly available from the Electron Microscopy Public Image Archive (EMPIAR) under accession codes EMPIAR-10017, EMPIAR-10028 and EMPIAR-10420. 

\section*{Code Availability}
The code implementing the proposed B factor estimation will be made publicly available upon publication.

\section*{Acknowledgements}

A.S. was supported in part by NSF DMS 2510039 and the Simons Foundation Math+X Investigator Award. 

The authors acknowledge the Ariel HPC Center at Ariel University for providing computing resources that have contributed to the research results reported in this paper.

\bibliographystyle{unsrt}
\bibliography{micrograph_based_b_factor_estimation}
\end{document}